\documentclass[aps,prl,twocolumn,amsmath,amssymb]{revtex4}
\usepackage{graphicx}
\usepackage{amsmath}

\begin{document}

\title{Spatially Modulated Interaction Induced Bound States and Scattering Resonances}
\author{Ran Qi and Hui Zhai}
\affiliation{Institute for Advanced Study, Tsinghua University, Beijing, 100084, China}
\date{\today}
\begin{abstract}
We study the two-body problem with a spatially modulated interaction potential using a two-channel model, in which the inter-channel coupling is provided by an optical standing wave and its strength modulates periodically in space. As the modulation amplitudes increases, there will appear a sequence of bound states. Part of them will cause divergence of the effective scattering length, defined through the phase shift in the asymptotic behavior of scattering states. We also discuss how the local scattering length, defined through short-range behavior of scattering states, modulates spatially in different regimes. These results provide a theoretical guideline for new control technique in cold atom toolbox, in particular, for alkali-earth-(like) atoms where the inelastic loss is small.

\end{abstract}
\maketitle

Feshbach resonances (FR) and optical lattices (OL) are two major techniques in cold atom toolbox. FR can be used to control the interactions by tuning a bound state in the so-called ``closed channel" to the scattering threshold via magnetic field, laser field or external confinement \cite{RMP,OFR,CIR}. At resonance, the $s$-wave scattering length diverges and the system becomes a strongly interacting one. OL can strongly modify the single particle spectrum of atoms, which suppress the kinetic energy so that the interaction effects are enhanced. With these two methods, many interesting many-body physics, such as BEC-BCS crossover, superfluid to Mott insulator transition and strongly correlated quantum fluids in low dimensions, have been studied extensively in cold atom systems during the last decade \cite{review}.

In this letter we theoretically study a new control tool for cold atom system. It is analogous to OL because it also makes use of two counter propagating laser fields that lead to a periodic modulation of laser intensity in space; however, its main effect is not acted on single particle, but manifested on the interaction term when two particles collide with each other. It generates a spatial modulation of two-body interaction, i.e. the two-body interaction potential not only depends on the relative coordinate of two particles under collision, but also depends on their center-of-mass coordinate.  As far as we know, this is a situation not encountered in interacting systems studied before, ranging from high-energy and nuclear physics to condensed matter systems. One can expect a spatial modulated interaction potential will result in many fascinating phenomena.

\begin{figure}[tbp]
\includegraphics[height=1.5in, width=2.5 in]
{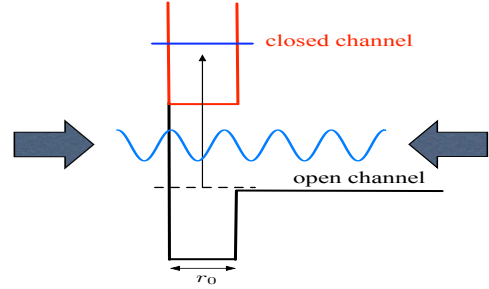}
\caption{A schematic of the model we studied. See text for detail description. \label{schematic}}
\end{figure}

The explicit model under consideration is schematically shown in Fig. \ref{schematic}. The open and closed channels are different orbital states and are coupled by laser field. Such an optical FR has been studied before for uniform laser intensity \cite{OFR}, and has been observed experimentally for both alkaline and alkaline-earth-like Yb atoms \cite{exp}. In this letter we shall consider the situation that the laser field is a standing wave whose intensity, and therefore the coupling strength between open and closed channel, is modulated periodically in space. Such a setup allows one to control spatial modulation of inter-atomic interaction on the scale of sub-micron. Recently, this has been realized in $^{174}$Yb condensation \cite{Kyoto}, although the optical standing wave is a pulsed one. Alkiline-earth(-like) atom like Yb is particular suitable for such an experiment because the narrow $^1S_0$-$^3P_1$ inter-combination transition line can avoid large inelastic scattering loss. In this experiment, a spatially modulated mean-field energy has been observed from diffraction pattern in a time-of-flight imagine \cite{Kyoto}. However, the theoretical study of this system is still very limited, and even the two-body problem has not been studied. In this work we show that surprises indeed arise even in the two-body problem of this model.

{\it Coupled Two-channel Model.} We consider a two-body Hamiltonian for a FR in which the open and closed channels are modeled by two square well potentials \cite{cheng}
\begin{equation}
\mathcal{H}=-\frac{\hbar^2}{4m}\nabla^2_{\bf {R}}-\frac{\hbar^2}{m}\nabla^2_{\bf {r}}+v({\bf R},{\bf r})\label{chin}
\end{equation}
where ${\bf R}=({\bf r_1}+{\bf r_2})/2$ and ${\bf r}={\bf r_1}-{\bf r_2}$. For $r<r_0$,
\begin{equation}
v({\bf R},{\bf r})=\left(\begin{array}{cc}-V_{\text{o}}&\hbar \Omega({\bf R}) \\ \hbar\Omega({\bf R}) & -V_{\text{c}}\end{array}\right)
\end{equation}
and for $r>r_0$,
\begin{equation}
v({\bf R},{\bf r})=\left(\begin{array}{cc}0 & 0 \\ 0 & +\infty \end{array}\right).
\end{equation}
In this model $V_{\text{o}}$ is given by the background scattering length $a_{\text{bg}}$ as $\tan(k_{\text{o}}r_0)/(k_{\text{o}}r_0)=1-a_{\text{bg}}/r_0$ where $k_\text{o}=\sqrt{mV_\text{o}}/\hbar$, and $V_{\text{c}}$ is determined by the binding energy of closed channel molecule $\epsilon_\text{c}$ through $V_\text{c}=\hbar^2\pi^2/(mr^2_0)-\epsilon_\text{c}$. The size of inter-atomic potential $r_0$ is much smaller than all the other length scales. Conventionally, the inter-channel coupling $\Omega({\bf R})$ is a constant independent of ${\bf R}$. Such a model captures all key features of a FR \cite{RMP,cheng}. A bound state appears at threshold and causes scattering resonance at $\Omega_0=\sqrt{\epsilon_\text{c}/|\beta|}$, and the scattering length is given by \cite{cheng}
\begin{equation}
a_\text{s}=a_\text{bg}\left(1-\frac{\beta\Omega^2}{\epsilon_\text{c}+\beta\Omega^2}\right)  \label{uniform_a}
\end{equation}
where $\beta=32 r_0 a_\text{bg}/(9\pi^2)$.

\begin{figure}[tbp]
\includegraphics[height=1.4in, width=3.0 in]
{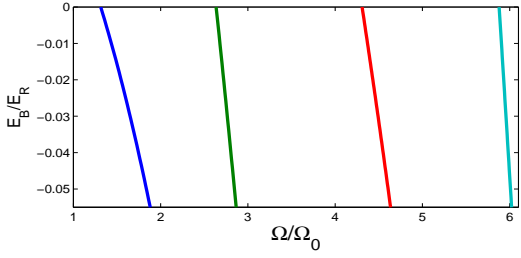}
\caption{First four bound states as a function of the amplitude of coupling $\Omega$. The parameters for this plot are $q=0$, $\epsilon_\text{c}=0.05 E_\text{R}$, $Ka_\text{bg}=-0.01$ and $Kr_0=10^{-3}$. $E_\text{R}=\hbar^2 K^2/m$ is taken as energy unit. \label{bound_state}}
\end{figure}

Now consider the situation $\Omega$ depends on ${\bf R}$. Solving the Schr\"odinger equation follows two steps: (i) in the regime $r<r_0$, because the $-\hbar^2\nabla^2_{{\bf r}}/m$ term commutes with the Hamiltonian, we consider the wave function of following form
\begin{equation}
\psi_\text{o}=\frac{\sin kr}{r}a({\bf R}); \   \  \psi_\text{c}=\frac{\sin kr}{r}b({\bf R})
\end{equation}
where $a({\bf R})$ and $b({\bf R})$ satisfy a coupled equation
\begin{eqnarray}
\left[-\frac{\hbar^2}{4m}\nabla^2_{\bf R}-V_\text{o}\right]a({\bf R})+\Omega({\bf R})b({\bf R})=\epsilon a({\bf R})\\
\left[-\frac{\hbar^2}{4m}\nabla^2_{\bf R}-V_\text{c}\right]b({\bf R})+\Omega({\bf R})a({\bf R})=\epsilon b({\bf R})
\end{eqnarray}
where $\epsilon=E-\hbar^2 k^2/m$. There will be a set of eigen-function $a_{l}({\bf R})$, $b_l({\bf R})$ and $k_l$ that give rise to the same energy $E$. The eigen wave function in the regime $r<r_0$ should be assumed as
\begin{equation}
\psi({\bf R},{\bf r})=\sum\limits_{l}A_l \frac{\sin k_l r}{r}\left(\begin{array}{c} a_l({\bf R}) \\ b_l({\bf R}) \end{array}\right)
\end{equation}
(ii) The superposition coefficient $A_l$, the binding energy $E$ for bound states, as well as the phase shift $\delta(E)$ for scattering states, are determined by matching the wave function in the regime of $r>r_0$ at $r=r_0$ for any ${\bf R}$.

Hereafter we will consider an explicit situation where $\Omega({\bf R})=\Omega\cos (K x)$ ($x$ denotes the $x$-component of ${\bf R}$). Note that there is still a discrete translation symmetry $x\rightarrow x+2\pi/K$, we can introduce a good quantum number ``crystal momentum" $q$.  In the regime $r>r_0$, $\psi_\text{c}=0$, and for the bound states whose energy $E<\hbar^2 q^2/(4m)$, $\psi^q_{\text{o}}(x,r)$ can always be expanded as
\begin{equation}
\psi^q_\text{o}(x,r)=e^{iqx}\sum\limits_{n}U^q_{n}e^{inKx}\frac{e^{-r\sqrt{(q+nK)^2-4m E/\hbar^2}}}{r}\label{Bloch}
\end{equation}
Eq. (\ref{Bloch}) can be viewed as the Bloch wave function for molecules. And for the low energy scattering state whose energy is greater than but close to $\hbar^2 q^2/(4m)$, we have
\begin{align}
\psi^q_\text{o}(x,r)&=e^{iqx}\left(U_0\frac{\sin(kr-\delta)}{r\sin\delta}\right.\nonumber\\
&\left.+\sum\limits_{n\neq 0}U^q_{n}e^{inKx}\frac{e^{-r\sqrt{(q+nK)^2-4m E/\hbar^2}}}{r}\right)\label{scattering}
\end{align}
where $k=\sqrt{mE/\hbar^2-q^2/4}$, and $\delta$ is a function of $k$.

\begin{figure}[tbp]
\includegraphics[height=1.5in, width=3.0 in]
{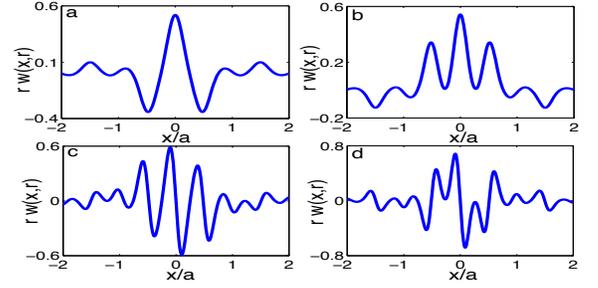}
\caption{$rw(x,r)$ (where $w(x,r)$ is the ``wannier wave function") for the first four bound states. $a=2\pi/K$ is the ``lattice spacing". The parameters for this plot are $\epsilon_\text{c}=0.05 E_\text{R}$, $Ka_\text{bg}=-0.01$, $Kr_0=10^{-3}$ and $Kr=0.1$.  \label{wannier}}
\end{figure}

 {\it Results 1-- Bound States:} In contrast to the uniform case where there is only one bound state when $\Omega>\Omega_0$, in this case we find a sequence of bound states as $\Omega$ increases, as shown in Fig. \ref{bound_state}. This is because the periodic structure of coupling $\Omega({\bf R})$ leads to a `` band structure " for the molecules, and as the coupling strength increases, the molecules with zero crystal momentum but in different bands touch the scattering threshold one after the other. We can introduce the ``wannier" wave function as
 \begin{equation}
 w(x-x_0,r)=\int_{-K/2}^{K/2}e^{iqx_0}\psi^q_\text{o}(x,r)dq
 \end{equation}
As shown in Fig. \ref{wannier}, the ``wannier" function for the bound states that appear at larger $\Omega$ have more oscillation, which means that they come from higher bands. This can also be illustrated from the symmetry of $U_n$ in the Bloch function of Eq. (\ref{Bloch}), as summarized in the Table \ref{symmetry} for the first four bound states. The first two bound state has even parity while the other two have odd parity.

\begin{figure}[tbp]
\includegraphics[height=2.0in, width=3.2 in]
{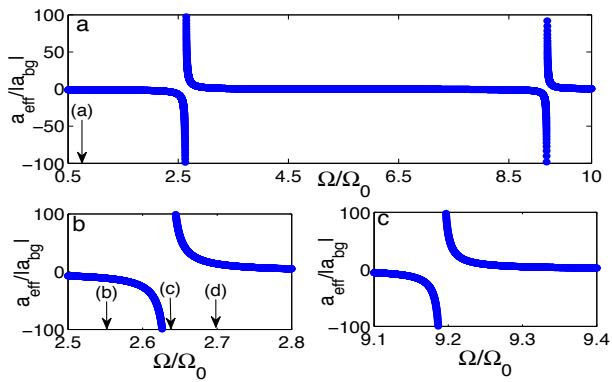}
\caption{The effective scattering length defined as Eq. (\ref{aeff}) $a_\text{eff}/|a_\text{bg}|$ as a function of $\Omega/\Omega_0$. (b) and (c) are enlarged plot around $\Omega/\Omega_0=2.64$ (b), $9.20$ (c). The arrows indicate the positions at which we plot the local scattering length $a_\text{loc}$ in Fig. \ref{Fig-aloc}(a-d). \label{Fig-aeff}}
\end{figure}

\begin{table}[tbp]
\begin{ruledtabular}
\begin{tabular}{|c|c|c|c|c|c|}
 & $U_{-2}$ & $U_{-1}$& $U_0$ & $U_1$ & $U_2$ \\
\hline
 1st & 0 & $+$ & 0 & $+$ & 0\\
\hline
2nd & $+$ & 0 & $+$ & 0 & $+$ \\
\hline
3rd & $+$ & 0& 0& 0& $-$ \\
\hline
4th & 0 & $+$ & 0 & $-$ & 0\\
\end{tabular}
\caption{Symmetry of Bloch wave function for the first four bound states \label{symmetry}}
\end{ruledtabular}
\end{table}

{\it Results 2 -- Effective Scattering Length:} For the scattering state wave function, at large ${\bf r}$ only the first term in Eq. (\ref{scattering}) will not exponentially decay, and the asymptotic behavior of the scattering wave function is still the same as that in the uniform case. Hence we can introduce an effective scattering length as
\begin{equation}
a_\text{eff}=\lim_{k\rightarrow 0}\frac{\tan\delta(k)}{k}.\label{aeff}
\end{equation}
Note that though the interaction is spatially dependent, the effective scattering length defined as Eq. (\ref{aeff}) is a spatial independent one. Among the first four bound states, $a_\text{eff}$ only diverges when the second bound state appears at threshold, as one can see by comparing Fig. \ref{Fig-aeff}(a) with Fig. \ref{bound_state}. This is because the divergence of $a_\text{eff}$ implies the first term in Eq. (\ref{scattering}) goes like $1/r$, which should be smoothly connected to a zero-energy bound state with non-zero $U_0$. Therefore, for the other three bound states whose $U_0=0$, their coupling to the low-energy scattering states vanish and will not cause divergency of $a_\text{eff}$. In Fig. \ref{Fig-aeff}(c) we show that $a_\text{eff}$ diverges when the sixth bound state (whose $U_0\neq 0$) appears at scattering threshold, but the width of resonance becomes narrower compared to Fig. \ref{Fig-aeff}(b) because this bound state comes from higher band and its coupling to low-energy scattering state ( i.e. the absolute value of $U_0$) is smaller.

\begin{figure}[tbp]
\includegraphics[height=2.0in, width=3.3 in]
{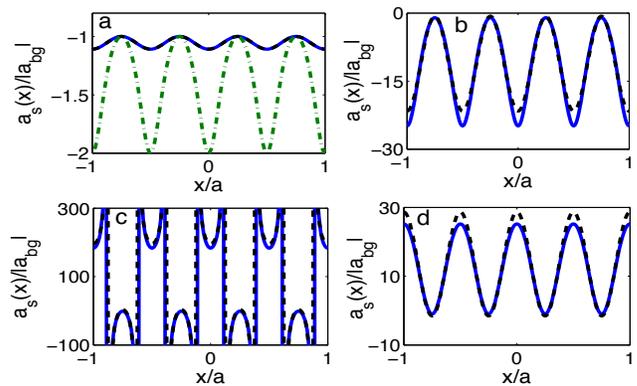}
\caption{The local scattering length $a_\text{loc}$ as a function of position $x/a$ for $\Omega/\Omega_0=0.71, 2.55, 2.64$ and $2.7$ (a-d). The solid blue line is calculated results, the black dashed line is the fitting formula Eq(\ref{approx_a}) or (\ref{approx_b}), and the green dash-dotted line in (a) is from simple replacement formula Eq. (\ref{naive}).   \label{Fig-aloc}}
\end{figure}

{\it Results 3 -- Local Scattering Length:} At short distance the wave function Eq. (\ref{scattering}) satisfies the Bethe-Peierls contact condition and display $1/r-1/a_{\text{loc}}(x)$ behavior, hence we can introduce a local scattering length as
\begin{equation}
a_{\text{loc}}(x)=-\lim\limits_{r\rightarrow r_{0}}\frac{ r\psi_\text{o}(x,r)}{\partial_{r}(r\psi_\text{o}(x,r))}\label{aloc}
\end{equation}
Unlike in the uniform case, $a_\text{eff}$ and $a_{\text{loc}}$ are different. Similar situation has also been encountered for scattering in confined geometry \cite{CIR}, lattices \cite{Cui} and mixed dimension \cite{tan}. What is unique here is that $a_\text{loc}$ is spatially dependent. Naively, one may think that $a_\text{loc}(x)$ can be obtained by replacing $\Omega$ in Eq. (\ref{uniform_a}) by local $\Omega(x)$, i.e.
\begin{align}
a_\text{loc}(x)&=a_\text{bg}\left(1-\frac{\beta\Omega^2\cos^2(Kx)}{\epsilon_\text{c}+\beta\Omega^2\cos^2(Kx)}\right)\label{naive2} \\
&\approx a_\text{bg}\left[1-\beta\Omega^2\cos^2(Kx)/\epsilon_\text{c}\right] \label{naive}
\end{align}
where the second line is valid for small $\Omega$. This formula in fact corresponds to an oversimplified approximation in our model that the kinetic energy term of the center-of-mass motion ($-\hbar^2\nabla^2_{\bf {R}}/(4m)$) is completely ignored in Eq. (\ref{chin}). In fact, what we really obtained from the wave function Eq. (\ref{scattering}) is
\begin{align}
a_\text{loc}(x)&=\frac{1-\sum_{m\neq 0}U_{m}\cos(mKx)/U_0}{a^{-1}_{\text{eff}}-\sum_{m\neq 0}U_m |m|K\cos(mKx)/(2U_0)}\nonumber \\
&\approx \frac{1-2U_{2}\cos(2Kx)/U_0}{a^{-1}_{\text{eff}}-2U_2 K\cos(2Kx)/U_0}
\label{alocal}
\end{align}
The second line is also valid when $\Omega$ is not too large, so the coefficient $U_{m>2}$ is small enough that can be ignored.

Away from a resonance, $Ka_\text{eff}\ll 1$, Eq. (\ref{alocal}) can be well approximated as
\begin{equation}
a_\text{loc}(x)=a_\text{eff}\left[1-\frac{2U_2}{U_0}\cos(2Kx)\right] \label{approx_a}
\end{equation}
In fact, we show in Fig. \ref{Fig-aloc}(a), (b) and (d) that the formula Eq. (\ref{approx_a}) (dashed black line) is a very good approximation to the actual results (solid blue line). In Fig. \ref{Fig-aloc}(a) we show the simple replacement formula Eq. (\ref{naive2}) already significantly deviates from the actual results in weak coupling regime. From Fig. \ref{Fig-aloc}(b) and (d) one can also see that the mean value of $a_\text{loc}(x)$ changes sign as $a_\text{eff}$ changes sign. At resonance, $a^{-1}_\text{eff}\rightarrow 0$, Eq. (\ref{alocal}) can be approximated as
\begin{equation}
a_\text{loc}(x)=\frac{1}{K}\left[1-\frac{U_0}{2U_2\cos(2Kx)}\right]\label{approx_b}
\end{equation}
We show in Fig. \ref{Fig-aloc}(c) that Eq. (\ref{approx_b}) is also a very good approximation to actual $a_\text{loc}$ at resonance. Hence, we show that $a_\text{loc}$ behaves very differently in the regime nearby or away from a scattering resonance.

{\it Implications to Many-body Physics:} In summary, we have revealed a number of novel features in the two-body problem with a spatially modulated interaction potential, which have strong implications for many-body physics and provide new insights for developing new tools for quantum control in cold atom systems.

First, when $a_\text{eff}$ diverges, the system enters a strongly interacting regime and is expected to exhibit universal behavior, which can even be manifested in the high temperature regime \cite{high-T}. For a two-component Fermi gas, it provides a new route toward BEC-BCS crossover physics, and ``high-temperature" superfluid may exist in this regime. The periodic structure will add new ingredient to the crossover physics.

Secondly, for the low-energy states whose energy $|E|\ll E_\text{R}$, the energy dependence of scattering length can be ignored and the many-body system can be effectively described by a pseudo-potential model:
\begin{equation}
\hat{H}=-\sum\limits_i\frac{\hbar^2\nabla^2_{\bf {r}_i}}{2m}+\sum\limits_{ij}\frac{4\pi \hbar^2a_{\text{loc}}({\bf R}_{ij})}{m}\delta^{3}(\mathbf{r_{ij}})\frac{\partial}{\partial r_{ij}}r_{ij},\label{single}
\end{equation}
where ${\bf R}_{ij}=({\bf r}_i+{\bf r}_j)/2$ and $r_{ij}={\bf r}_i-{\bf r}_j$. It is very important that $a_\text{loc}({\bf R})$ in the pseudo-potential of Eq. (\ref{single}) is given by Eq. (\ref{alocal}) from the two-body calculation, so that a two-body problem of the Hamiltonian Eq. (\ref{single}) can produce correct low-energy eigen-wave function and the effective scattering length as from model potential.

For bosons, with a mean-field approximation, Eq. (\ref{single}) implies that the interaction energy should take the form 
\begin{equation}
E_{\text{mf}}=\frac{4\pi\hbar^2}{m}\int a_{\text{loc}}(x)n^{2}(x) dx\label{Eint}
\end{equation}
which leads to a modulation of condensate density $n(x)$ and self-trapping nearby the minimum of $a_\text{loc}(x)$. It is very likely a strong enough modulation of condensate density will eventually result in the loss of superfluidity and the system enters an insulating phase. If so, it provides a completely different mechanism for superfluid to insulator transition where the transition is not driven by suppression of kinetic energy as in conventional OL.

{\it Final Comments:} In this work we choose a coupled two square-well model whose advantage is that the physics can be demonstrated in a simple and transparent way. However, some more sophisticated effects in real system, such as the inelastic loss, are ignored. We have also implemented more systematic scattering theory which includes these effects and found that the physics discussed here will remain qualitatively unchanged. These results will be published elsewhere \cite{Qiran}.

Moreover, the formalism used in this work can be easily generalized to other realizations of spatial modulation of interactions. For instance, in a magnetic FR, one can consider the presence of a magnetic field gradient so that the closed channel molecular energy varies spatially. This effect is particularly important for a narrow resonance. One can also optically couple the closed channel molecule to another molecular state via a bound-bound transition, which leads to a periodic variation of molecule energy \cite{Rempe}. Similar effects as discussed in {\it Results 1-3} also present in these cases \cite{Qiran}.

{\it Acknowledgements.} We thank Xiaoling Cui, Zeng-Qiang Yu, Peng Zhang and Zhenhua Yu for helpful discussions. This work is supported by Tsinghua University Initiative Scientific Research Program, NSFC under Grant No. 11004118 and NKBRSFC under Grant No. 2011CB921500.

{\bf Appendix:} In this appendix, we present some details of solving the two-body Schr\"odinger equation.  Using the discrete translation symmetry, we expand 
\begin{align}
a^q(x)=e^{iqx}\sum_{n}e^{inKx}a^{q}_n\\
b^q(x)=e^{iqx}\sum_{n}e^{inKx}b^{q}_n,
\end{align}
$a_n$ and $b_n$ satisfy coupled matrix equation
\begin{eqnarray}
\left(\frac{\hbar^2(q+nK)^2}{4m}-V_\text{o}\right)a^q_n+\frac{\Omega}{2}\left(b^q_{n-1}+b^q_{n+1}\right)=\epsilon^q a^q_n\\
\left(\frac{\hbar^2(q+nK)^2}{4m}-V_\text{c}\right)b^q_n+\frac{\Omega}{2}\left(a^q_{n-1}+a^q_{n+1}\right)=\epsilon^q b^q_n
\end{eqnarray}
This matrix has a set of eigen-values $\epsilon^q_l$ and their eigen-vectors $\{a^q_{l,n},b^q_{l,n}\}$. Hence there are a set of wave functions sharing the same energy $E$, which ar
\begin{eqnarray}
\psi^q_{\text{o},l}=e^{iqx}\frac{\sin (k^q_l r)}{r}\sum\limits_{n}e^{inKx}a^{q}_{l,n}\\
\psi^q_{\text{c},l}=e^{iqx}\frac{\sin (k^q_l r)}{r}\sum\limits_{n}e^{inKx}b^q_{l,n}
\end{eqnarray}
where $k^q_l=\sqrt{m(E-\epsilon^q_l)}/\hbar$ is a function of $E$. In general, the eigen-states take the form
\begin{equation}
\psi_q=\sum\limits_{l}A^q_{l}\left(\begin{array}{c}\psi^q_{\text{o},l} \\\psi^q_{\text{c},l}\end{array}\right)=e^{iqx}\sum\limits_{n}e^{inKx}\left(\begin{array}{c}\varphi^q_{\text{o},n} \\\varphi^q_{\text{c},n}\end{array}\right)
\end{equation}
where
\begin{eqnarray}
\varphi^q_{\text{o},n}({\bf r})=\sum\limits_{l}A^q_l\frac{\sin(k^q_l r)}{r}a^q_{l,n}; \  \ \varphi^q_{\text{c},n}({\bf r})=\sum\limits_{l}A^q_l\frac{\sin(k^q_l r)}{r}b^q_{l,n} \nonumber
\end{eqnarray}

For bound states, to match the boundary condition with Eq. (\ref{Bloch}) at $r_0$ in both open and closed channels, we obtain a matrix equation $M^q_{kl}(E)A^q_l=0$ where
\begin{align}
&M^q_{2n+1,l}=\sin(k^q_l r_0)b^q_{l,n}\nonumber\\
&M^q_{2n,l}=(k^q_l\cos(k^q_l r_0)+\sqrt{ (q+nK)^2-\frac{4mE}{\hbar^2}}\sin(k^q_l r_0))a^q_{l,n}\nonumber
\end{align}
Therefore for a given $q$ the eigen-energy $E$ is determined by $\text{Det}(M)=0$ and 
\begin{equation}
U^q_{n}=e^{r_0\sqrt{\frac{\hbar^2(q+nK)^2}{4m}-E}}\sum\limits_{l}A_l a^{l}_n \sin(k^l r_0)\label{Uqn}
\end{equation}
For the scattering states, $M^q_{2n+1,l}$ and $M^q_{2n,l}$ ($n\neq 0$) are the same as bound state, while 
\begin{eqnarray}
M^q_{0,l}=\sin(k^q_{l}r_0)\cos(k r_0-\delta)k-\cos(k^q_{l}r_0)\sin(k r_0-\delta)a^{l}_0\nonumber
\end{eqnarray}
where $k=\sqrt{mE/\hbar^2-q^2/4}$. In this case $\text{Det}(M)=0$ gives rise to the relation between phase shift $\delta$ and energy $E$. $U^q_n$ ($n\neq 0$) is also the same as Eq. (\ref{Uqn}), while for $n=0$, 
\begin{equation}
U^q_0=\frac{\sin\delta}{\sin(k r_0-\delta)}\sum\limits_{l}A_l \sin(k^q_l r_0)a^l_{0}
\end{equation}


\begin{thebibliography}{99}

\bibitem{RMP}
C. Chin, R. Grimm, P. Julienne, and E. Tiesinga, Rev. Mod. Phys. {\bf 82}, 1225 (2010).

\bibitem{OFR}
P. O. Fedichev, Yu. Kagan, G. V. Shlyapnikov, and J. T. M. Walraven, Phys. Rev. Lett. {\bf 77}, 2913 (1996); J. L. Bohn and P. S. Julienne, Phys. Rev. A {\bf 56}, 1486 (1997).

\bibitem{CIR}
M. Olshanii, Phys. Rev. Lett. {\bf 81}, 938 (1998); T. Bergeman, M. G. Moore, and M. Olshanii, Phys. Rev. Lett. {\bf 91}, 163201 (2003).

\bibitem{review}
I. Bloch, J. Dalibard, and W. Zwerger, Rev. Mod. Phys. {\bf 80}, 885 (2008).

\bibitem{exp}
F. K. Fatemi, K. M. Jones, and P. D. Lett, Phys. Rev. Lett. {\bf 85}, 4462Ð4465 (2000);
M. Theis, {\it et al.}
Phys. Rev. Lett. {\bf 93}, 123001 (2004) and K. Enomoto, K. Kasa, M. Kitagawa, and Y. Takahashi,
Phys. Rev. Lett. {\bf 101}, 203201 (2008).

\bibitem{Kyoto}
R Yamazaki, S. Taie, S. Sugawa, and Y. Takahashi, Phys. Rev. Lett. {\bf 105}, 050405 (2010).

\bibitem{cheng}
C. Chin, arXiv: 0506313.

\bibitem{Cui}
X. Cui, Y. Wang, and F. Zhou, Phys. Rev. Lett. {\bf 104}, 153201 (2010) and H. P. B\"uchler, Phys. Rev. Lett. {\bf 104}, 090402 (2010)

\bibitem{tan}
Y. Nishida and S. Tan, Phys. Rev. Lett. {\bf 101}, 170401 (2008)

\bibitem{high-T}
T. L. Ho and E. J. Mueller, Phys. Rev. Lett. {\bf 92}, 160404 (2004).

\bibitem{Qiran}
R. Qi, P. Zhang and H. Zhai in preparation.

\bibitem{Rempe}
D. M. Bauer, {\it et al.}
Nature Physics {\bf 5}, 339 (2009).

\end{thebibliography}
\end{document}